\documentclass[twocolumn,prd,showpacs,floatfix]{revtex4}
\usepackage{graphicx}
\usepackage{dcolumn}
\usepackage{bm}
\begin{document}

\preprint{IFT/12/02}

\title{ Special relativity constraints on
the effective constituent theory of hybrids}

\author{Stanis{\l}aw D. G{\l}azek}
\affiliation{ Institute of Theoretical Physics, Warsaw University, 
              ul.  Ho{\.z}a 69, 00-681 Warsaw, Poland }
\author{Adam P. Szczepaniak}
\affiliation{Department of Physics and Nuclear Theory Center,
Indiana University, Bloomington, Indiana 47405}

\date{\today}

\begin{abstract}
We consider a simplified constituent model for relativistic
strong-interaction decays of hybrid mesons. The model is  constructed
using rules of renormalization group procedure  for effective
particles in light-front quantum field theory,  which enables us to
introduce low-energy phenomenological parameters. Boost covariance is
kinematical and special  relativity constraints are reduced to the
requirements of  rotational symmetry. For a hybrid meson decaying into
two  mesons through dissociation of a constituent gluon into  a
quark-anti-quark pair, the simplified constituent model  leads to a
rotationally symmetric decay amplitude if the  hybrid meson state is
made of a constituent gluon and a  quark-anti-quark pair of size
several times smaller than  the distance between the gluon and the
pair, as if the pair originated from one gluon in a gluonium state in
the same effective theory.
\end{abstract}

\pacs{13.25.Jx, 11.80.-m, 13.90.+i}

\maketitle

\section{ Introduction }
\label{sec:i}
Hybrid mesons, in particular those with exotic quantum numbers, will
play important role in the phenomenology of confinement and studies of
low energy gluons. For example, decay patterns of gluonic excitations
in  hybrid mesons are expected to be quite different from those that
characterize  ordinary meson resonances. It is predicted that exotic  
meson decays are dominated by the so-called ``S+P'' channels that contain 
pairs of mesons, one with the valence quark-anti-quark pair in the 
$S$-wave relative motion, and the other with a pair in  a $P$-wave, 
{\it e.g.}, $\pi b_1$, $\pi f_1$, or $\pi a_1$~\cite{isgur}.  
In contrast, non-exotic meson decays are dominated
by ``S+S'' channels,  such as $\pi\pi$, $\rho\rho$, $K{\bar K}$, {\it
etc.} These predictions are based on a simple, non-relativistic
constituent-quark picture.  However, the lightest exotic mesons are
expected to be as heavy as  $1.6-2$ GeV and their decays will produce
mesons that move  relativistically. Therefore, this article addresses
the issue of  constraints that special relativity imposes on the
constituent picture  of such decays. The other basic  requirement 
satisfied by the relativistic approach developed here is that in principle 
it is derivable from quantum field theory in  agreement with rules of 
renormalization group for low energy  parameters in the constituent 
Hamiltonians.

In order to analyze data from experiments where photons
are absorbed on a hadronic target and produce intermediate hadronic
states with masses on the order of 1.5-2.5 GeV, which subsequently
decay into mesons $\pi$, $\rho$, or $b$, one must construct a viable
picture  of strongly interacting bound states. These states,
especially  $\pi$-mesons, move with velocities comparable to the speed
of light, and  the whole photo-production and decay process cannot be
described using  wave functions for hadrons at rest, or in slow
motion. At the same time,  the principles of quantum field theory have
to be obeyed precisely  if one demands a clear connection between a
phenomenological constituent picture of the process and QCD. Lessons
from QED are of great value but QCD  interactions are much stronger
and pose additional problems with  non-perturbative and relativistic
effects, which QED is not telling us  how to solve beyond Feynman's
perturbation theory.

So far, theoretical insights into the exotic mesons structure are
provided  by lattice simulations \cite{lat1}, and Coulomb-gauge QCD
\cite{CG}, in  addition to model studies \cite{PP1,PP2}. It follows
from these studies  that the lowest mass and spin exotic mesons are
most probably dominated by a constituent gluon coupled to a
constituent  quark-anti-quark pair. In a more complete picture, one
would like the  exotic candidate states to be obtainable, at least in
principle, from a  precisely defined relativistic Hamiltonian for
QCD. The decay process  should be described by the same
Hamiltonian. To satisfy these requirements,  one would have to derive
the effective constituent particles from quantum  field theory. The
derivation would have to include description of the  constituents'
spin and angular momenta, and how rotating and boosting of  bound
states to high velocities is accomplished. Such comprehensive
relativistic constituent picture is not yet established in QCD and
hadronic  constituents are far from being understood as well as the
non-relativistic  constituents of atoms are in QED. However, there
exist reasonable theoretical grounds for hope that the required
relativistic description of effective  quarks and gluons can be found
using renormalized light-front (LF) QCD  Hamiltonians in $A^+=0$
gauge. The reasons for the hope go beyond the  pioneering works of
Lepage and Brodsky \cite{LB} and Wilson et al. \cite{W6},  and include
results that suggest that the effective constituents are  derivable in
a perturbative renormalization group calculus for the  corresponding
creation  and annihilation operators \cite{GQCD}. The same calculus is
also suggested  to be able to produce all generators of the Poincar\'e
algebra in the form ready to use in the effective-particle Fock-space
basis \cite{SGTM}. If  these generators were available, the desired
construction of hadrons,  including exotic hybrid states and fast
moving mesons, could be carried  out precisely enough to make specific
predictions for partial waves branching  ratios in decay amplitudes,
and the QCD constituent theory of hybrids could  be put to definite
tests.

The LF approach is a potential alternative to the standard form of
Hamiltonian dynamics, where states evolve in usual time. There,
rotational symmetry is kinematical, but one needs to worry about
boosting  the constituent wave functions for the fast moving bound
states and their constituents \cite{RQM}. Nevertheless, much progress
has been achieved in building  the constituent representation in the
usual Coulomb-gauge quantization in the rest frame \cite{CG,CG2}. The
connection between bare field quanta and  quasi-particles,
corresponding to a constituent basis, was obtained and  related to
spontaneous chiral symmetry breaking. Furthermore, in the gluonic
sector, the Coulomb-gauge canonical quantization reproduces the the
lattice  gauge theory results for the spectrum of the low-lying gluon
modes, {\it e.g.},  glueballs, and the excited heavy-quark adiabatic
potential \cite{lat1}. Still, the Lorentz boost covariance of
observables in the instant formulation  in Minkowski space is not yet
established, and poses problems analogous to  rotational symmetry
issue in the boost invariant LF scheme. Both are major  theoretical
challenges. Here, we will concentrate on the rotational symmetry  in
the LF description of hadrons in terms of their constituents.

Note, however, that the LF QCD path is not sufficiently developed for
immediate identification of all important features of the effective
constituent dynamics in hadronic processes. The theory requires
initial  guidance from phenomenological analysis to begin scanning
through and  cataloging of the infinite set of potentially relevant
operators that  are allowed to appear in the effective QCD
Hamiltonian. In view of the  magnitude of the task, before one dives
into the full complexity of LF QCD,  one should first find out if
rotational symmetry of observables has any chance  to be obtained in
the boost-invariant picture of hadronic states made of a  small number
of effective-particle Fock-components. The common belief is that a
physically adequate relativistic scheme cannot be achieved using  a
small number of degrees of freedom. The structural integrity
requirement  for the effective-particle LF-scheme thus implies that
the rotational  symmetry condition should be possible to satisfy with
some natural easiness, visible already in the most elementary versions
of the scheme. Therefore, the  question of how hard it is to build a
rotationally symmetric model of a  hybrid decay amplitude with a
minimal number of constituents needs to be  answered before a search
for approximate solutions to quantum field theory  in terms of
effective constituents is pursued on a larger scale. Here, we  report
results of the feasibility test for this scheme, which were obtained
with all possible simplifications one could afford preserving only
the most basic and universal constraints of special relativity, quantum
field theory,  and renormalization group for effective particles. All
details beyond the core features are removed for transparency of the
bottom-line structure.

In principle, the dynamics of hadronic constituents is determined by
the  effective-particle Hamiltonian. Once such a Hamiltonian is
constructed, one  can attempt to solve the corresponding Schr\"odinger
equation and calculate the hybrid decay in terms of the
effective-particle  interactions and states.  The basic study reported
here is carried out instead by simplifying the initial  canonical
Hamiltonian theory to the ultimate minimum, deriving the effective
theory in a crudest lowest-order approximation, and using model wave
functions typical for phenomenology instead of self-consistent
solutions   of the currently unknown effective theory. Thus, in
practice, we  calculate a transition amplitude for a hand-made one
three-effective-particle bound state to decay into two hand-made
two-effective-particle bound states, and we check if the result has
any chance to respect rotational symmetry.

This paper is organized as follows. Section \ref{sec:key} describes
the  key elements of the constituent scheme used here. It is assumed
that the interaction  responsible for the hybrid decay comes from an
effective Hamiltonian  obtained from a local quantum field theory via
renormalization group  procedure for effective particles. Section
\ref{sec:swf} describes details  of the model states of hybrids and
mesons, and the transition amplitude for  the hybrid decay into the
two mesons. Wave functions of the incoming and  outgoing states are
written down on the basis of well-known phenomenology, but without
solving the effective Schr\"odinger equation. Nevertheless, we allow a
whole set of free parameters in them in order  to cover a large range
of possible structures and find out if it is possible (and if so, then
for what choices of the parameters) to obtain rotationally symmetric
results. Numerical results from our review of  the angle dependence of
the decay amplitude are discussed in Section \ref{sec:nr}. Section
\ref{sec:c} draws conclusions that concern phenomenological studies of
hybrid production and decay processes and can be directly useful in
comparing data from future experiments with a theory of effective
particles in QCD.

\section{ Key elements }
\label{sec:key}

In order to remove the basic issues of spin, gauge invariance,
current conservation, and effective mass of gluons, from the preliminary
constituent study, and to focus on general quantum mechanical and
relativistic features of the effective particle dynamics, one can
assume that gluons are scalars. In order to do better than that at
the current level of understanding of the effective gluon dynamics,
one would have to clearly answer questions concerning Lorentz
transformation properties of massive gluons with spin one and only
two, instead of three polarization states. Since the answer is
lacking, the issue is avoided by the arbitrary simplification.

With the gluon spin issue removed, the scalar-iso-scalar bare gluon can
still  be coupled to an iso-doublet of bare quarks like in a Yukawa
theory. Thus, the  bare ``gluon'' has only color as in QCD, while the
coupling to bare quarks is  given by the term
\begin{equation}
 {\cal H}_{int} = g \bar \psi A^a t^a \psi  \quad ,
\label{bareint}
\end{equation}
where the gluon field does not have a vector index. In this model, the
first   order solution of renormalization group equations for the
effective gluon  coupling  to effective quarks produces the term
\cite{SGMW}
\begin{equation}
g f_\lambda \bar \psi_\lambda A^a_\lambda t^a \psi_\lambda \quad ,
\label{effint} 
\end{equation}
where $\lambda$ denotes the renormalization group parameter. It has
dimension of mass. Its scale can be intuitively associated with the
transverse size of the effective particles and $f_\lambda$ denotes
the vertex form factor. The width of the form factor in momentum space
is given by $\lambda$. The effective particles that can be
considered  appropriate for the lowest Fock-sector picture of hadrons
are expected to emerge when $\lambda$ is lowered by solving similarity
renormalization group equations for Hamiltonians down to the scale of
hadronic masses, possibly to the values on the order of 1 to several 
GeV, when one aims at description of a hybrid meson in terms of a 
quark-anti-quark pair and a gluon.

The factor $g$ stands for the vertex strength function $v_\lambda$
that at some value of the momentum transfer defines a running
coupling constant in the Hamiltonian \cite{GQCD}, denoted by $g_\lambda$. 
In the first order solution of renormalization group equations, 
$v_\lambda$ equals $g$. In higher orders, this coupling would be 
replaced by $g_\lambda$ and the latter could be quite large in QCD 
when $\lambda$ is of the order of hadronic masses. However, the 
large coupling is not sufficient to make the effective interaction 
strong enough to wash out the constituent picture of the individual 
hadrons, as it would happen in a local canonical theory with equally 
large coupling constants. The reason is that the vertex form factor 
cuts off interactions with momentum transfers larger than $\lambda$, 
and they are effectively much weaker, like in nuclear physics. Therefore, 
the bound state picture with a small number of massive constituents may 
be still valid when $\lambda$ is small. Eq.(\ref{effint}) provides also a 
potentially valid structure for at least a part of the interaction 
responsible for the decay of a model constituent ``gluon'' into the 
effective quark and anti-quark pair.  The ``gluon'' decay factor in the 
model hybrid decay amplitude can thus be assumed to be given by the 
product of the form
\begin{equation}
g f_\lambda(12) \bar u_1 v_2 \quad , \label{gdf}
\end{equation}
times the corresponding color factor $\chi^\dagger_{c1} t^a
\chi_{c2}$.  One may simplify the constituent picture even farther for
the purpose of this study, and replace fermionic quarks by scalars,
too. The required  interaction, analogous to (\ref{bareint}), would be
\begin{equation}
 H_{int} = g \xi^\dagger A^a t^a \xi  \quad ,
\label{bareints}
\end{equation}
where $\xi$ is the colored scalar quark field, and the interaction
factor  in the decay, analogous to (\ref{gdf}), would reduce to
\begin{equation}
g f_\lambda(12) \quad , \label{gds}
\end{equation}
the color factor being the same as for fermions. Both cases of
spin-1/2 and spin-0 constituents are described in this paper to show
that the features they display are general and independent of
spin. This is one  of the reasons why one can expect them to occur
also in the future QCD  calculations.

Even a modest attempt at a self-consistent dynamical solution of a
relativistic field-theoretic model with coupled three-body and
two-body bound states is a large project. Our goal here is not to
carry out such project but to find out if the proposed scheme has   a
chance to produce rotationally symmetric decay amplitude. Moreover,
even if a constituent picture may satisfy the special relativity
constraints, the correct effective QCD theory is not known yet.  In
order to see how strong the relativity constraints are, the simplest
thing to do is to assume that the outgoing mesons and decaying hybrid
are both scalars and iso-scalars, and to write down candidates for
their  wave functions in the effective constituent basis with {\it a
priori}   unknown parameters that are constrained only by reasonable
guesses about  their natural sizes. If the rotational symmetry
constraints cannot be satisfied within this class of models, it is
unlikely that a simple  constituent model can ever be built in more
complex cases with additional  dynamical constraints. Thus, for
example, we assume here only that a meson  has a Gaussian
momentum-space relative-motion wave function with width  $\beta$.  Its
spatial radius is of the order of $1/\beta$ (if $\beta$ is not too
large in comparison to the constituent masses and the meson mass since
relativistic effects may change connection between $\beta$ and the
radius making it less constraining \cite{KS}). Thus, in the first
approximation, a meson of size 0.5 fm is expected to have $\beta \sim
0.4$ GeV. In the theory of scalar effective constituents, no further
considerations are required for constructing plausible test candidates
for the relative-motion wave functions, and none is incorporated here.

In the more complex case of fermions, one has to decide in addition
how to  combine spins in the wave functions for effective
constituents. Solutions  of effective fermion theory are also not
known, and one is currently in  the similar position here as in trying
to guess hadronic wave functions  in phenomenological studies in the
nonrelativistic quark model~\cite{Isgur1}.  The simplest standard
phenomenological procedure is to assume that the spin  wave functions
for the lowest effective Fock components are the same as for
multiplets of Poincar\'e group representations for free particles,
{\it i.e.}, products of free Dirac spinors in the case of
fermions. For  scalar (colorless) hadrons made of $q \bar q$ pairs,
one would use spin  factors of the form $\bar u_1 v_2$ (the color
factor would be  $\chi^\dagger_{c1} \chi_{c2}$). The spin amplitudes
would be multiplied  by the relative momentum-dependent factors that
would be of similar form as in the scalar case.

The relative-motion scalar wave-function factors represent binding and
allow for shifting the total free energy of the intermediate
constituent  particles away from that of the energy shell value for
the hybrid itself  and the outgoing meson products. This shift in
energy is associated with conservation of the total three-momentum of
all particles.  In the LF Hamiltonian dynamics, however, the conserved
momenta are not the same as in the dynamics developed in Minkowski's 
time. Instead, the conserved momentum components (in conventional 
notation) are $p^+=p^0 + p^3$ and $p^\perp$, where $\perp =$ 1, or 2. 
The role of  energy is played by $p^-$, which for a free particle of 
mass $m$ would  be $p^- = (p^{\perp \, 2} + m^2)/p^+$.

The key Lorentz invariance test, associated with the complicated
dynamics of rotations in the LF coordinates, is reduced to the
question if the off-shell  energy shift in the direction of  $p^-$
does, or does not introduce  significant artificial angular
asymmetries in the model hybrid decay  amplitude. For example, can one
avoid a sizable dependence of the amplitude on the angles at which the
scalar mesons fly away from the scalar hybrid decay event in the
hybrid center  of mass system of reference? If such angle dependence
would unavoidably develop  in the simplest scalar or fermion model to
intolerable degree, one would have  to doubt that the simple
constituent picture could suddenly become valid in  QCD. If rotational
symmetry can be approximately obtained, one could hope that
proceeding along the line outlined in Refs. \cite{SGTM} and 
\cite{SGMW}, corrections can be refined and eventually a precise 
theory obtained.

It was mentioned earlier in Section \ref{sec:i} that when seeking
Hamiltonian  dynamics for effective particles in the instant (or usual
time) form, one would  have to make tests of boost invariance. One
could then say that the deviations  from rotational symmetry one may
observe in a LF approach provide a measure of  how large deviations
from boost covariance can be expected in models based on  the standard
form of field-theoretic constituent dynamics, even if they respect 
rotational symmetry in some frame of reference exactly.

\section{ The model}
\label{sec:swf}
The $q{\bar q}$-meson and $q{\bar q}g$-hybrid wave functions are 
constructed by analogy with decomposing a product of representations of
the Lorentz group for two and three noninteracting particles into 
irreducible representations defined by their invariant mass and 
$J^{PC}$ quantum numbers. The free-particle on-shell invariant 
mass constraint corresponds to a state with a wave function proportional 
to a $\delta$-function in the invariant mass. In the model, mesons 
are constructed by superposing such states in the form of an integral 
with a Gaussian wave-function factor that depends on the invariant 
mass \cite{Lor}, or, equivalently, on the length of the corresponding
center-off-mass momenta. Before a theoretically sound, realistic 
constituent Hamiltonian in QCD is identified, such approach to building 
models of the non-perturbative hadron structure is quite common, 
{\it e.g.}, in the constituent models, or, with some variation, in 
QCD sum rules. 
\subsection{$q{\bar q}$ Mesons}
\label{subsec:mesons}
The two-constituent (or $q{\bar q}$) meson state is written  as
\begin{equation}
|p \rangle = \sum_{12} \int[12] 
\,\, p^+ \tilde \delta(1+2 -p) \Psi_{J^{PC}}(1,2)
\,\, b^\dagger_1 d^\dagger_2 |0\rangle \; ,
\label{p}
\end{equation}
where $1$ and $2$ collectively represent all quantum numbers of the
quark and anti-quark, {\it i.e.} the LF 3-momenta, $p^+$, $p^\perp$, 
spins, and charges, respectively. The (free) Lorentz-invariant 
integration measure, denoted in Eq. (\ref{p}) and similarly later 
by $[12]$, is given by 
\begin{eqnarray}
{dk^+_1 d^2k^\perp_1 \over 16\pi^3 k^+_1}
{dk^+_2 d^2k^\perp_2 \over 16\pi^3 k^+_2} 
= 
{dx_{12} d^2 k^\perp_{12} \over 16\pi^3 x_{12}(1-x_{12})} 
{dP^+_{12} dP_{12}^\perp  \over 16\pi^3 P^+_{12}} \, ,
\end{eqnarray}
where for all three components of the 3-momenta, $P_{12} = k_1 + k_2$, 
$k_{12} = (1-x_{12})k_1-x_{12}k_2$, and $x_{12} = k^+_1/P^+_{12}$.
Also, $\tilde \delta(1+2-p) = 16\pi^3 \delta(P^+_{12}-p^+)
\delta^2(P^{\perp}_{12}-p^\perp)$. Due to the kinematical 
nature of Lorentz boost invariance of the LF scheme, even in the 
fully interacting theory, the coordinates $P^{+,\perp}_{12}$, $x_{12}$, 
and $k^\perp_{12}$, enable one to separate the relative constituent 
motion from the bound-state center-of-mass motion. In general, in 
the $n$-particle case, the relative-motion variables that are 
independent of the bound-state momentum $p$ are given by 
($i=1,\cdots, n$)
\begin{equation}
x_i \equiv k^+_i/p^+, \;\; 
k^\perp_i \equiv k^\perp_i - x_i p^\perp \, . 
\label{jak}
\end{equation}
 
For given quantum numbers $J^{PC}$, the wave function 
$\Psi_{J^{PC}}(1,2)$ is a product of color, flavor, spin, and 
orbital parts, and can be written as, 
\begin{equation}
\Psi_{J^{PC}}(1,2) = \chi^\dagger_{c_1}C\chi_{c_2}
\,\, \chi^\dagger_{i_1}I\chi_{i_2}
\,\, \chi^\dagger_{s_1}S(1,2)\chi_{s_2}
\,\, \psi_p(1,2) \,\, ,
\end{equation}
with $C = 1/\sqrt{3}$ for a color singlet, $I = 1/\sqrt{2}$ 
for an iso-scalar, and $I=\vec \tau/\sqrt{2}$ for an iso-vector 
meson. Since iso-spin is irrelevant for our studies of Lorentz 
symmetry, further explicit consideration of the model is limited 
to iso-scalar meson states.

The model spin-wave function $\chi^{\dag}S(1,2)\chi$ originates from 
the product of free Dirac spinors. In this analysis, however, we 
do not perform a detailed phenomenological study for all exotic meson
decay channels and we are not concerned about particulars of the
spin-orbit couplings for the corresponding $q\bar q$ states. For example, 
$J^{PC} = 0^{-+}$, $q{\bar q}$-meson wave function could involve the 
product ${\bar u}_1 \gamma_5 v_2$ as the spin factor. However, for the 
critical check if the rotational symmetry can be obtained at all, 
we do not need to insist on constructing states with arbitrary 
$J^{PC}$ quantum numbers for the mesons in question (here, two, or 
three, non-interacting constituents). Instead, we choose a simplest 
scalar form ${\bar u}_1 v_2$ for all decay products. This ansatz 
already introduces some spin-orbit structure and we can use it to 
check the sensitivity of our numerical estimates to the presence of 
relativistic spin-orbit couplings. Note that in the calculation of 
decay amplitude the spins of the constituents are summed over and it 
does not matter what spin basis on chooses. Nevertheless, the natural 
choice in the LF scheme is to use Melosh spinors, 
\begin{eqnarray}
u(k\sigma) = B(k,m_q) u(0\sigma) \;\; , \nonumber \\
v(k\sigma) = B(k,m_q) v(0\sigma) \;\; , \nonumber \\
v(k\sigma) = C u^*(k\sigma) = i\gamma^2 u^*(k\sigma) \;\; ,
\end{eqnarray}
with 
\begin{equation}
B(k,m) = {1\over \sqrt{k^+ m}}
  [ \Lambda_+ k^+ + \Lambda_- (m + k^\perp \alpha^\perp)] ,
\end{equation}
which represents a LF-symmetry boost from rest to momentum 
$k$ for a free particle of mass $m$, using conventions 
$\Lambda_\pm = {1\over 2}\gamma_0 \gamma^\pm$, 
$u^{T}(0\sigma) =  \sqrt{2m_q}[\chi^T_\sigma,0]$, 
$v^{T}(0\sigma) = \sqrt{2m_q} [0,\xi^T_{-\sigma}]$, 
and $\xi_{-\sigma}=-i\sigma^2\chi_\sigma$.
Adopting these conventions, one obtains 
\begin{equation}
S(1,2) =  {1\over  \sqrt{x_{12}(1-x_{12})}}
                   \left[m_q(2x_{12} - 1)\sigma^3 + 
                   k_{12}^\perp \sigma^\perp\right] i \sigma^2 .
\end{equation}

The orbital wave function $\psi_p(1,2) \equiv \psi_p(\vec k_{12})$ 
is chosen to be Gaussian, {\it i.e.}, 
\begin{equation}
\psi_p(\vec k_{12}) = N_p \,\, \sqrt{{\cal M}_{12}\over 2 m_q} 
\,\, \sqrt{1 \over 2( {\cal M}_{12}^2 - 4 m_q^2)}
\,\, \exp{\left[{-{\vec k}_{12}^{\, 2} \over 2 \beta_p^2}\right]}
\label{eq:psi}
\end{equation}
where the 3-vector $\vec k_{12}$ has the $\perp$ components 
equal to $k_{12}$ and the $z$-component is defined by \cite{DN}, 
\begin{equation}
{\cal M}^2_{12} = (k_1 + k_2)^2  = 
{ m^2_q + k_{12}^{\perp \, 2} \over x_{12}(1-x_{12})}= 4({\vec k}^{\,
  2}_{12} + m^2_q) \; ,
\end{equation}
so that $k^3_{12} = (x_{12}-1/2){\cal M}_{12}$.
Note that the subscript $p$ in $\psi_p(\vec k_{12})$ is not supposed
to mean that $\psi_p(\vec k_{12})$ depends on the bound state momentum
$p$, but rather that the wave function corresponds to the meson with
quantum numbers collectively denoted by $p$, and the notational 
simplification should not be a source of confusion when we
consider three mesons simultaneously.
  
In terms of the 3-vector $\vec k_{12}$, the integration measure 
over the relative motion of the two constituents, is given by 
\begin{equation}
{dx_{12} d^2 k^\perp_{12} \over 16\pi^3 x_{12}(1-x_{12})} =
{4 d^3 \vec k_{12} \over 16 \pi^3 {\cal M}_{12}} \; .
\end{equation}

The normalization condition $\langle p|p'\rangle = 
p^+ \tilde \delta(p-p')$,  implies, 
\begin{equation}
 {1\over 16 \pi^3}
\int { 4 d^3 k_{12} \over {\cal M}_{12}} 
2( {\cal M}_{12}^2 - 4 m_q^2) |\psi_p(1,2)|^2 =1 \; ,
\end{equation}
and
\begin{equation}
N_p = {2 \pi^{3/4} \over \beta_p}\sqrt{ 2 m_q \over \beta_p} \; . 
\end{equation}
When we consider scalar constituents, the spin factor
$2( {\cal M}_{12}^2 - 4 m_q^2)$ is absent and the second square-root
factor in Eq. (\ref{eq:psi}) is removed, so that the normalization 
constant remains the same. 
\subsection{Hybrid Meson}
\label{subsec:hybrid} 
The construction of the hybrid meson state is analogous. 
\begin{equation}
|h\rangle = \sum_{123} \int[123] 
\, h^+ \tilde \delta(1+2+3 -h)
\Psi_{J^{PC}}(1,2,3) \, b^\dagger_1 d^\dagger_2 a^\dagger_3
|0\rangle ,
\end{equation}
where the color, flavor, spin, and orbital wave-function 
factors are given by 
\begin{eqnarray}
\Psi_{J^{PC}}(1,2,3) & = & 
\,\, \chi^\dagger_{c_1}C_h^{c_3}\chi_{c_2}
\,\, \chi^\dagger_{i_1}I_h\chi_{i_2}
\,\, \chi^\dagger_{s_1}S_h(1,2)\chi_{s_2} \nonumber \\
& \times & \psi_h(1,2,3) \; , 
\end{eqnarray}
with $C^{c_3}_h = t^{c_3}/2$, $I_h = 1/\sqrt{2}$, and  the
quark-anti-quark spin wave function is again of the from 
${\bar u}_1 v_2$, so that 
\begin{equation}
S_h(1,2)  =  {1-x_3 \over \sqrt{x_1 x_2}}
                   \left[m_q{x_1 - x_2 \over 1 - x_3}\sigma^3 + 
                   k_q^\perp \sigma^\perp\right] i \sigma^2.
\end{equation}
The kinematical momentum variables satisfy Eq. (\ref{jak}) 
with $p^+,p^\perp$ replaced with the hybrid LF-3-momentum, $h$. 
Similarly, the 3-particle integration measure is given by, 
\begin{equation}
\prod_{i=1}^3 {{dk^+_i d^2k^\perp_i} \over {16\pi^3 k^+_i}} 
= 
{ 4 d^3 k_q \over 16 \pi^3 {\cal M}_q}
{d^3 k_g {\cal M}_{qg} \over 16 \pi^3 
\sqrt{ m_g^2 + k_g^2} \, \sqrt{{\cal M}_q^2 + k_g^2}} \; ,
\end{equation}
where the relative momentum 3-vectors ${\vec k}_q = 
(k^\perp_q,k^3_q)$ and 
${\vec k}_g = (k^\perp_g,k^3_g)$, are defined by
\begin{eqnarray}
k^\perp_q = (x_2 k_{\perp 1} - x_1 k_{\perp 2})/(1 - x_3) \; , 
\nonumber \\
k^3_q = \left({x_1\over 1-x_3} - {1 \over 2}\right) {\cal M}_q \; ,
\end{eqnarray}
and 
\begin{eqnarray}
k^\perp_g = (1-x_3) k^\perp_3 - x_3 (k^\perp_1 + k^\perp_2) \; ,
\nonumber \\
k^3_g = 
{x_3^2({\cal M}^2_q + k^{\perp \, 2}_g)-(1-x_3)^2(m^2_g +
k^{\perp \, 2}_g) \over  2 x_3 (1-x_3) {\cal M}_{qg}} \; .
\end{eqnarray}
${\cal M}_q$ and ${\cal M}_{qg}$ are the quark-anti-quark and
quark-anti-quark-gluon invariant masses, 
\begin{equation}
{\cal M}^2_q = (k_1 + k_2)^2 = 
(1-x_3)^2 { m^2_q + k_q^{\perp \, 2} \over x_1 x_2} \; ,
\end{equation}
\begin{eqnarray}
{\cal M}^2_{qg} & = & (k_1 + k_2 + k_3)^2  \nonumber \\
 & = & {m_q^2 + k_1^{\perp \, 2} \over x_1} + 
       {m_q^2 + k_2^{\perp \, 2} \over x_2} + 
       {m_g^2 + k_3^{\perp \, 2} \over x_3} - 
       h^{\perp \,2} \nonumber \\
 & = & {{\cal M}_q^2 + k_g^{\perp \, 2} \over 1-x_3} + 
       {m_g^2 + k_g^{\perp \, 2} \over x_3} \; .
\end{eqnarray}

In terms of the relative momenta, the 3-particle orbital
wave function $\psi_h(1,2,3) \equiv \psi_h(\vec k_q, \vec k_g)$ 
is assumed to be
\begin{eqnarray}
\psi_h(\vec k_q, \vec k_g) & = & N_h \,\, \sqrt{{\cal M}_q\over 2 m_q}
   \,\, \sqrt{2m_q + m_g \over {\cal M}_{qg}} \nonumber \\
& \times &
   \,\, \sqrt{ \sqrt{m_g^2 + {\vec k}^{\,2}_g} \over m_g}
   \,\, \sqrt{ \sqrt{ {\cal M}_q^2 + {\vec k}^{\,2}_g} \over 2 m_q}
 \,\, {1 \over \sqrt{8 {\vec k}_q^{\, 2}}} \nonumber \\
& \times &
 \,\, \exp{\left[{-{\vec k}_q^{\, 2} \over 2 \beta_{hq}^2}\right]} 
 \,\, \exp{\left[{-{\vec k}_g^{\, 2} \over 2 \beta_{hg}^2}\right]} \; .
\end{eqnarray}
Its structure follows from the requirement that it is a product
of two Gaussian functions of the momenta $\vec k_q$ and $\vec k_g$, 
with additional factors chosen so that the wave function normalization 
condition has a nonrelativistic appearance of a six-dimensional integral
over the momenta. In the normalization condition 
$\langle h|h'\rangle = h^+ \tilde \delta(h-h')$,  
fermion spinors introduce a factor, which is compensated
by the factor $(8 {\vec k}_q^{\, 2})^{-1/2}$ in the wave
function. This factor is dropped when we consider scalar constituents. 
So, in both cases of the fermion and scalar constituents 
the normalization constant is the same and equals
\begin{equation}
N_h = {2 \pi^{3/4} \over \beta_{hq}}\sqrt{ 2 m_q \over \beta_{hq}} \,\,
      {2\pi^{3/4} \over \beta_{hg}}             
 \sqrt{ {8 m_q m_g \over 2 m_q + m_g} \, {1\over \beta_{hg}} } \; .
\end{equation}
\subsection{Decay}
\label{subsec:decay}
The constituent-gluon decay into a constituent quark-anti-quark
pair is assumed to be driven by the Hamiltonian interaction term 
obtained from first-order solution to the renormalization group 
equations for effective particle Hamiltonians in quantum field 
theory. In QCD, the decay amplitude of a hybrid meson into two 
mesons, calculated in first order in this interaction term, has 
the form
\begin{equation} 
\langle p b| H_{I \, QCD \, \lambda} |h\rangle = 
\tilde \delta (p+b-h) \,\, {\cal A}(p,b,h) \quad ,
\end{equation}
where the interaction term is given by \cite{GQCD}
\begin{eqnarray}
 H_{I \, QCD \, \lambda} = \sum_{123}\int [123] \,\, \tilde \delta (1+2-3) 
\,\, g \,\, f_\lambda(1,2,3) \nonumber \\
\,\, \chi^\dagger_{c_1}t^{c_3}\chi_{c_2}
\,\, \chi^\dagger_{i_1}\chi_{i_2}
\,\, \chi^\dagger_{s_1}S_{QCD}(1,2,3) \chi_{s_2} 
\,\, b_1^\dagger d_2^\dagger a_3 ,
\end{eqnarray}
The spin-factor $\chi^\dagger_{s_1}S_{QCD}(1,2,3)\chi_{s_2}$ equals
$\bar u_1 \not\! \varepsilon_3 v_2$ and $\varepsilon_3$ denotes the
constituent-gluon polarization four-vector. The form factor $f_\lambda$ 
is given by 
\begin{equation}
f_\lambda (1,2,3) = f_\lambda({\cal M}^2_{q}-m^2_{g0})= 
\exp{\left[-{\cal M}^4_{q}/\lambda^4\right]} \; ,
\end{equation}
since the effective gluon mass $m_{g0}$ is zero in the first-order 
$H_{QCD \,\lambda}$. 

In our elementary study of the Poincar\'e symmetry constraints on the 
constituent picture, the spin factor $\chi^\dagger_{s_1}S_{QCD}(1,2,3)
\chi_{s_2}$ is replaced by $\bar u_1 v_2$, which one would obtain in 
the same scheme in Yukawa theory \cite{SGMW}, and, using 
$\kappa^\perp_{12} = x_2 k^\perp_1 - x_1 k^\perp_2$, $S_{QCD}(1,2,3)$ 
is set equal to 
\begin{equation}
S(1,2) = {1\over \sqrt{x_1 x_2}}
\left[m_q(x_1 - x_2)\sigma^3 + \kappa^\perp_{12})
\sigma^\perp\right] i \sigma^2 \quad \; ,
\end{equation}
{\it i.e.}, we remove the gluon spin effects by assuming it is
a spin-0 particle instead. 

For convenience of normalization of the coupling 
constant in the test calculation, $g$ is written as 
$g = g_\lambda v_\lambda$ and  $v_\lambda = 
\exp{\left[16 m_q^4/\lambda^4\right]}$, so that for 
a pair produced at the threshold given by the constituent
quark masses, $ g f_\lambda = g_\lambda $, which is kept 
the same in all cases, corresponding to $\alpha_\lambda =  g^2_\lambda/
(4\pi) = 1$. 

\begin{figure}[htb]
\includegraphics[scale=0.4]{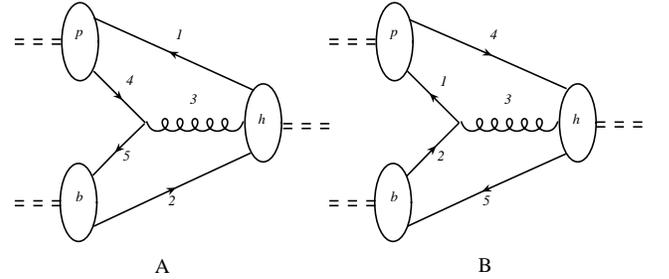}
\caption{\label{fig:1}
 Hybrid meson decay amplitude into two non-exotic mesons $p$ (such as 
 $\pi$-meson) and  $b$ (such as $b_1$-meson).}
\end{figure}

In the decay, both meson states, $|p\rangle$ and $|b\rangle$
in the product $|pb\rangle = |p\rangle |b\rangle$, are constructed 
in the same way and may differ only trough the meson mass parameter, 
$m_p$ or $m_b$, and the width of the orbital wave-function $\psi$,
$\beta_p$ or $\beta_b$. Thus, the decay amplitude is a sum of two terms, 
\begin{equation}
 {\cal A}(p,b,h) = A(p,b,h) + B(p,b,h) \; ,
\end{equation}
that are schematically shown in Fig. \ref{fig:1}. 
The result of summing over all quantum numbers except momenta can be 
written in the term $A(p,b,h)$ as
\begin{eqnarray}
& &
\sum_{c_3} Tr\left[ C^\dagger_p   C^{c_3}_h   C_b^\dagger    t^{c_3}   \right] 
Tr\left[ I^\dagger_p         I_h   I_b^\dagger              \right] \nonumber \\
& &
Tr\left[ S^\dagger_p(1,4) S_h(1,2) S^\dagger_b(5,2) S_{QCD}(5,4)\right] \; ,
\end{eqnarray}
and in $B(p,b,h)$ as
\begin{eqnarray}
& &
\sum_{c_3} Tr\left[ C^\dagger_p  t^{c_3}         C_b^\dagger C^{c_3}_h   \right] 
Tr\left[ I^\dagger_p                  I_b^\dagger       I_h   \right] \nonumber \\
& &
Tr\left[ S^\dagger_p(1,4) S_{QCD}(1,2) S^\dagger_b(5,2)S_h(5,4)\right] \; . 
\end{eqnarray}
The full expression for the amplitude including the momentum
integrals, is given in the Appendix. 
\section{Numerical results}
\label{sec:nr}
This Section describes how the decay amplitude
$\cal A$ depends on the angle $\theta$ between the direction of 
flight of one of the mesons (the meson $p$) and the arbitrarily 
chosen $z$-axis, in the hybrid rest frame of reference, and how 
this dependence changes when the parameters of the model are 
varied. A true solution of QCD would not have this much freedom of 
varying parameters. However, if the model were found unable to 
satisfy the requirement of rotational symmetry with reasonable 
accuracy for any choice of the parameters introduced here, our
effective LF constituent picture with only lowest Fock sectors 
could be dismissed as a good approximation to start with.

\begin{figure}[htb]
\includegraphics[scale=0.4]{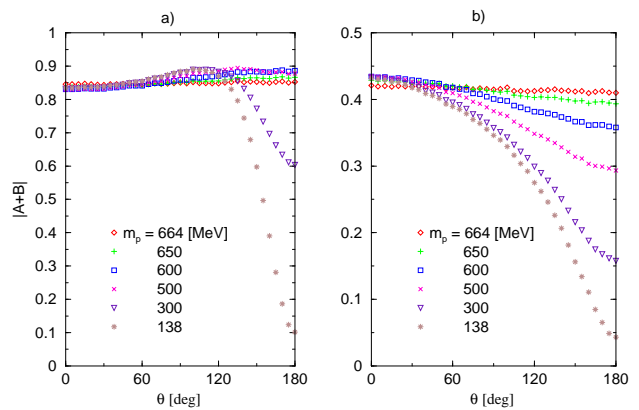}
\caption{\label{fig:2}
Angle dependence of the model hybrid decay amplitude $\cal A$ 
for various choices of the light meson mass, given in MeV as indicated. 
The smaller is the meson mass, the more important become relativistic 
effects and the larger is the violation of rotational symmetry. The 
left plot a) is for scalar constituents, and the right plot b) for 
fermion quarks. All points correspond to the same standard set of 
parameters from the first column of Table \ref{table:1}, labeled 
2a,b, and $m_p$ is varied from 664 MeV down to 138 MeV. See 
the text for details.} 
\end{figure}

Since the decaying hybrid and the outgoing mesons are constructed 
to be scalars, one should obtain an amplitude that does not depend 
on $\theta$ (the LF approach respects rotational symmetry around
the $z$-axis and all amplitudes we consider are automatically constant
as functions of the azimuthal angle $\phi$). No dependence on $\theta$ 
is always obtained when the sum of masses of the outgoing mesons 
is smaller by only a few MeV than the mass of the hybrid. In this 
case, both outgoing mesons move with speeds that are small in comparison 
to the speed of light. When at least one of the mesons becomes light 
and the available mass defect forces it to move fast, the amplitude 
begins to deviate from a constant. 
This is illustrated in Fig.~\ref{fig:2}. In Fig.~\ref{fig:2}a, 
we illustrate what happens in the model with scalar constituents 
representing quarks, and Fig.~\ref{fig:2}b shows what happens in 
the version with quarks represented by fermions. The same convention 
of two versions, a) for scalar and b) for fermion constituents, is 
kept in all graphs that follow. 
Fig. \ref{fig:2} shows that in the case of one of the decay products being 
light, such as the $\pi$-meson, the  decay amplitude develops practically 
100 \% deviation from rotational symmetry. This is clearly not acceptable 
and invalidates the initial picture based on a conventional choice of the 
model parameters, given in the first  column of Table \ref{table:1}, 
labeled 2a,b. The reason is that when the light meson moves against 
the $z$-axis direction, its longitudinal momentum, 
\begin{equation}
p^+ \; = \; \sqrt{m_p^2 + |\vec p \, |^2 \, } \; + \; p^3 \; ,
\end{equation}
becomes small. As a result of the $+$-momentum conservation, the quarks 
that make up this light meson must carry small fractions of the parent, 
hybrid momentum $h^+$. This configuration is suppressed if one employs 
parameters that would be naively considered as valid in a nonrelativistic 
model. In that case, the gluon decay vertex is taken directly from quantum 
field theory without the renormalization group form factor $f_\lambda$. 

However, varying parameters of the model one can observe that the drastic 
violation of rotational symmetry is not always present. Since the number 
of mass parameters is 10, and varying all of them produces a huge number 
of combinations to investigate, one is faced with a problem of how to do a 
systematic test. The condition of obtaining an amplitude that does not 
depend on the angle $\theta$ (an infinite number of conditions) is so 
strong that it may easily be impossible to achieve a constant amplitude 
with a small set of parameters. It turns out, however, that one can obtain 
small variation with angle for selected sets of the parameters that are 
discussed below. 

\begin{figure}[htb]
\includegraphics[scale=0.4]{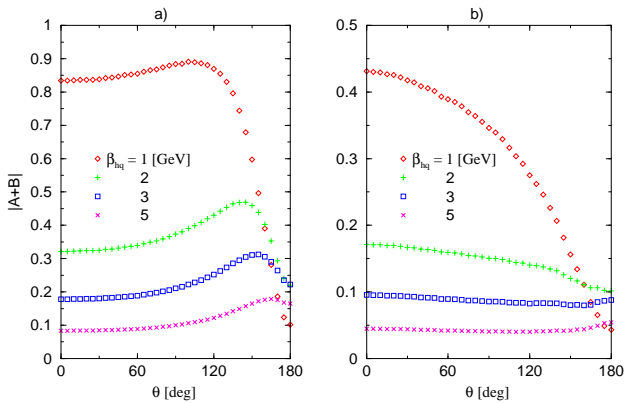}
\caption{\label{fig:3} 
Angle dependence of the decay amplitude $\cal A$ with $m_p = 0.1375$ 
MeV (corresponding to the lowest curves in Fig. \ref{fig:2}) for various 
choices of the quark-anti-quark subsystem size in the hybrid state. The 
size is indicated in terms of the momentum space wave function width 
$\beta_{hq}$ in units of GeV, as indicated (other parameters are constant 
and equal to the displayed in the second column of Table \ref{table:1},
labeled 3a,b). 
The spatially smaller is the pair sub-system the larger is the relative 
magnitude of the relativistic decay amplitude at large angles, where 
Fig. \ref{fig:2} displayed a deficiency. See the text for details. The 
left graph a) is for scalar constituents and the right graph b) for 
fermion quarks.}
\end{figure}

\begin{figure}[htb]
\includegraphics[scale=0.4]{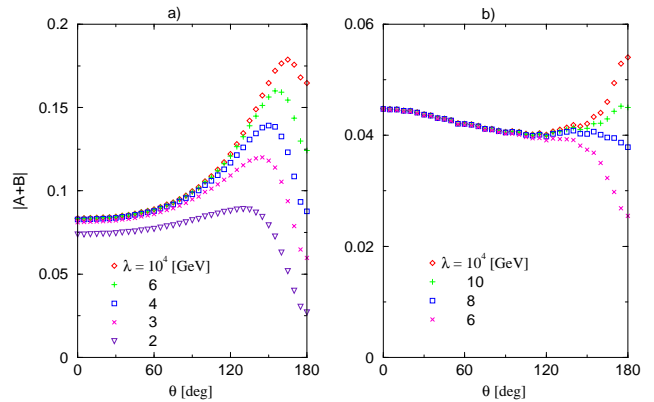}
\caption{\label{fig:4} 
The result of lowering the renormalization group form factor width
parameter, $\lambda$, given in GeV as indicated, in the cases with 
$m_p = 0.1375$ MeV and $\beta_{hq} = 5$ GeV (lowest curves in Fig. 
\ref{fig:3}). The smaller is $\lambda$ the more reduced is the amplitude 
at large angles. The effect confirms that lowering $\lambda$ may remove 
the large-angle increase in the amplitude due to large $\beta_{hq}$ 
independently of the constituent quark spin. These results suggest that 
the small size of the quark-anti-quark pair in the hybrid (large 
$\beta_{hq}$) may be correlated with the form factor width $\lambda$ 
in a relativistic LF-theory through rotational symmetry. The left graph a) 
is for scalar constituents and the right graph b) for fermion quarks.}
\end{figure}

\begin{figure}[htb]
\includegraphics[scale=0.5]{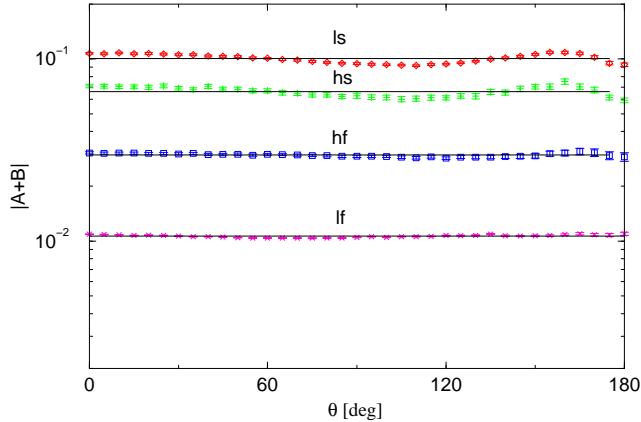}
\caption{\label{fig:5} 
Four cases of decay amplitudes $\cal A$ obtained by fitting wave-function 
parameters and $\lambda$ so that an amplitude has the least possible 
variation as a function of the angle (see columns 5hs, 5ls, 5hf, 5lf 
in Table \ref{table:1}). The two upper curves are obtained in the scalar 
model, and the two lower curves are obtained in the fermion model. In 
both models, the curves differ by the optimal choice of the quark mass, 
which is on the order of 160-170 MeV in the cases labeled as light, and 
365-375 MeV in the cases labeled as heavy. In all cases, $\beta_{hq} 
\sim \lambda \sim 4-5$ GeV, which suggests that the size of quark-anti-quark 
pair is related to the form-factor width as if the pair in the hybrid 
state were produced through splitting of one of two gluons in a gluonium 
state of the same effective theory. See details in the text.}
\end{figure}

The results are based on a systematic computer-added manual review of 
about 6000 cases and subsequent numerical minimization of the amplitude 
variation with respect to $\theta$. Since the six dimensional integrals 
were carried out using Monte Carlo integration (using the procedure 
{\it vegas} \cite{vegas}), the accuracy of the result of integration 
(standard deviation output from vegas) varied from a typical 0.1 - 0.3 \% 
to 2 \% in the worst cases, and the minimization of angle dependence 
(Powell's procedure \cite{Powell}) had to be carried out with a somewhat 
worse accuracy. Therefore, variations of the parameters that caused smaller 
variations of the amplitude $\cal A$ with $\theta$ than 1\% could not be 
kept under precise control. Thus, in the obtained sets of parameters, 
displayed in Table \ref{table:1} with three significant digits, the first 
is stable, while the second is highly probable and the third can be 
considered a result of statistical fluctuation. Figs. \ref{fig:3} and 
\ref{fig:4} describe the two dominant effects that we found scanning 
the space of parameters, and Fig. \ref{fig:5} shows results of the 
final searches for the local minima in the angle dependence. The error 
bars indicate the size of three standard deviations obtained in the 
Monte Carlo routine.

Since the decay of a hybrid into one meson as heavy as $b_1$ and 
another as light as $\pi$ is of key interest, the mass parameters 
$m_h = 1.9$ GeV, $m_b = 1.235$ GeV and $m_p = 0.1375$ GeV are kept
constant. The other seven parameters were first chosen to be (all in 
units of GeV, which will be omitted in the discussion) $m_q = 0.3$, 
$m_g = 0.8$, $\beta_p = \beta_b = \beta_{hg} = \beta_{hq} = 0.4$
and $\lambda = 4.0$. Then a code calculated $3^7$ cases (each for 
35 different values of $\theta$) with each of the 7 parameters changing 
through 3 values obtained by multiplying the initial values by 0.5, 1.0, 
and 2.0. The resulting cases were reviewed manually and additional 
calculations were performed to verify trends visible in the samples.
This way the two generic features illustrated on examples in Figs. 
\ref{fig:3} and \ref{fig:4} were identified. 

Fig. \ref{fig:3} shows two cases, a) for scalar, and b) for fermion 
quarks, of what happens with the decay amplitude when one makes the 
quark-anti-quark pair in the hybrid smaller and smaller in size in 
position space ($\beta_{hq}$ grows from 1 to 5 GeV). The amplitudes are 
lowered at small angles while they remain more or less stable around 
$\theta = 180^\circ $. This effect results from the fact that the 
larger domain of momenta of quarks in the hybrid is sampled by the
same momentum-space wave functions of mesons, and they see less of 
the total probability of finding the quark-anti-quark pair in the hybrid, 
while the large relative momentum-tail is sampled in a similar way. 
Thus, the deficiency observed in Fig. \ref{fig:2} at large angles can 
disappear if the quark-anti-quark pair in the hybrid is made small in its 
spatial size. This is understandable since quarks are allowed to move 
with momenta so large that they can easily have small longitudinal 
fractions $x$ required in the light meson, $p$, when it flies out 
against $z$-axis. However, in both cases a) and b) the amplitude 
overshoots for large $\theta$ and a flat case is not achieved. This 
is easier to see in Fig. \ref{fig:4}, which concerns the case of the 
lowest curves of $\beta_{hq} = 5.0$ GeV from Fig. \ref{fig:3}. 
Note that the remaining variation with angle is about thrice smaller 
in the case of fermions than in the scalar model, where it is still 
at the 100 \% level.

Fig. \ref{fig:4} shows another tendency in the parameter space that 
is related to a reduction of the overshooting due to large $\beta_{hq}$
visible in Fig. \ref{fig:3}. Namely, the pair of quarks that originate 
from the hybrid decay via gluon dissociation, have their relative momentum 
limited by the renormalization group form factor $f_\lambda$. In a bare 
quantum field theory, $\lambda = \infty$ and no such suppression is 
present. However, when one lowers $\lambda$, the gluon cannot be turned 
into a pair of very fast moving quarks through the effective interaction, 
and the amplitude is suppressed again. The interplay of $\beta_{hq}$ and 
$\lambda$ turns out to be the key ingredient of the model that at the 
end allows it to produce small deviations from a constant in the 
amplitude $\cal A$ as a function of $\theta$.

Detailed minimization of the $\theta$-dependence leads to the four 
cases shown in Fig. \ref{fig:5}. The deviation from constant was 
measured using the sum of squares of deviations from the average value,
or maximum of the modulus of deviations from the average. The values 
of masses, wave function widths, and form factor parameters that
produce the minimal angle variations, are given in Table \ref{table:1} 
in the last four columns. The remarkable feature is that the required 
$q \bar q$-wave function in the hybrid has roughly the same width as 
the form factor in the effective vertex and this width is systematically 
much larger than other width parameters (see Table \ref{table:1}). 
\begin{table}[htb]
\begin{ruledtabular}
\caption{\label{table:1} 
Parameters in Figs. \ref{fig:2}-\ref{fig:5} (in GeV). \# a is 
associated with scalar particles, and b with fermions. $^\dagger$ 
hs=heavy scalars, ls=light scalars, hf=heavy fermions, lf=light 
fermions. $^*$ The parameter is the same as in the left neighboring 
column. $^{**}$ The parameter is varying from curve to curve in the 
figure. }
\begin{tabular}{lllllllllll}
Fig. \#     &2a,b  &3a,b     &4a,b  &5hs$^\dagger$&5ls&5hf&5lf \\
$m_h$       &1.9   &$s^{*}$  &$s$   &$s$  &$s$  &$s$  &$s$  \\
$m_b$       &1.235 &$s$      &$s$   &$s$  &$s$  &$s$  &$s$  \\
$m_p$    &$v^{**}$ &0.1375   &$s$   &$s$  &$s$  &$s$  &$s$  \\
$m_q$       &0.3   &$s$      &$s$   &.365 &.175 &.375 &.161 \\
$m_g$       &0.8   &$s$      &$s$   &1.63 &2.46 &.867 &.845 \\
$\beta_p$   &0.4   &$s$      &$s$   &.375 &.363 &.180 &.142 \\
$\beta_b$   &0.4   &$s$      &$s$   &1.01 &.719 &.812 &.501 \\
$\beta_{hg}$&1.0   &$s$      &$s$   &.310 &.600 &.783 &.854 \\
$\beta_{hq}$&1.0   &$v$      &5.0   &4.37 &4.61 &4.50 &4.93 \\
$\lambda$   &10000 &$s$      &$v$   &4.44 &4.49 &4.50 &4.50 
\end{tabular}
\end{ruledtabular}
\end{table}

These results can be interpreted as that the requirement of rotational 
symmetry in our relativistic LF-constituent picture, forces the hybrid's
constituents to form a system made of a small-size quark-anti-quark-pair 
and a gluon, and the pair is about 5 to 10 times smaller than the 
distance between the gluon and the pair. In principle, this picture 
resembles a gluonium made of two gluons \cite{AP}, but with one of the 
gluons replaced by a $q\bar q$-pair. The observed correlation between 
$\beta_{hq}$ and $\lambda$, suggests that the $q \bar q$-pair can be 
seen to originate dynamically from a decay of one gluon in a gluonium, 
a picture perhaps derivable from a well-defined $H_{QCD \, \lambda}$. 

The amplitudes still depend on the other parameters than $\beta_{hq}$
and $\lambda$, and those can be varied in a considerable ranges causing 
residual changes in the angle dependence. The most pronounced effect 
is that flattest amplitudes are systematically arrived at when quark 
masses take only one of two values, about 160-170 MeV or about 370 MeV. 
Since in quantum field theory such as QCD one can associate the large 
renormalization group scale $\lambda$ with the smaller rather than the 
larger quark masses, it appears to be a remarkably consistent feature 
of the effective constituent approach that the lighter quark mass cases 
lead to amplitudes that vary less in $\theta$ than for the heavier quarks.
\section{Conclusion}
\label{sec:c}
The numerical study described in this work shows that constraints 
imposed  by requirements of rotational symmetry on parameters of a 
crude LF constituent model of a relativistic decay of a hybrid 
can be satisfied even if the model assumes that only lowest 
effective-particle Fock-space components are included in the wave 
functions of the participating hadrons. Since the boost symmetry 
is kinematical in the LF quantum field theory, and the model is 
built to be compatible with the rules of renormalized LF Hamiltonian 
dynamics for effective particles, it is interesting that a 
self-consistent picture of the hybrid state in such a simple class 
of models emerges without major obstacles as soon as one introduces 
the right width $\lambda$ of the effective vertex form factor. 
The key renormalization group aspect of the constituent model is 
that the special-relativity symmetry constraints are satisfied only
approximately. The accuracy is expected to improve when the 
renormalization group equations for effective constituent theory 
are solved including higher order terms than the first considered
here, and when the resulting constituent dynamics is solved precisely, 
which together would be analogous to restoring rotational symmetry 
in lattice gauge theory.

The physical picture of a hybrid meson that one can now attempt
to uncover by performing the detailed calculations in QCD, has the 
following spatial structure. A quark-anti-quark-pair of small
size and a gluon move around a common center, and this system of 
two objects is not far from a gluonium. In the latter case, one has 
a second gluon, instead of the small quark-anti-quark pair. 
Solution of the QCD Hamiltonian eigenvalue problem has a chance to 
be approximated this way only in terms of renormalized effective 
particles, for which the similarity renormalization group width 
parameter is comparable to the size of the wave-function parameters. 
The key difficulty of the calculation will be to properly include 
the constituent-gluon spin.
\begin{acknowledgments}
This work was supported by KBN grant No. 2 P03B 016 18 (SDG) and 
the US Department of Energy under contract DE-FG02-87ER40365 (APS). 
SDG would like to acknowledge the financial support and outstanding 
hospitality of the Nuclear Theory Center at Indiana University, 
where part of this work was done. SDG also thanks Mr. and Mrs. 
D. W. Elliott, of the Elliott Stone Company, Inc., Bedford, Indiana, 
for supplementary financial support. 
\end{acknowledgments}
\appendix* 
\section{}
In the model without gluon spin, one substitutes 
\begin{equation}
S_{QCD}(i,j) = S_p(i,j)=S_b(i,j)= S_h(i,j) = S(i,j)
\; , 
\end{equation}
and obtains 
\begin{widetext}
\begin{eqnarray}
&&{\cal A}(p,b,h) = (-1) {2\over 3} {1\over \sqrt{2}}  
{g_\lambda \over (16 \pi^3)^2} 
\int {dx_{14} d^2\kappa^\perp_{14} \over x_{14}(1-x_{14})}
\int {dx_{52} d^2\kappa^\perp_{52} \over x_{52}(1-x_{52})} 
{1\over x_3}  \nonumber \\ 
&& \times N_p N_b N_h \psi^*_p(1,4) \psi^*_b(5,2) 
T(1,2,4,5) \left[ A(1,2,4,5) + B(1,2,4,5) \right]  \nonumber \\
&& = -{16 \over 3} {g_\lambda \over (16 \pi^3)^2 } 
\int {d^3 k_b \over {\cal M}_b} 
\int {d^3 k_p \over {\cal M}_p} 
{1\over x_3} N_p N_b N_h \psi^*_p(\vec k_{14}) \psi^*_b(\vec k_{52}) 
T(1,2,4,5) \left[ A(1,2,4,5) + B(1,2,4,5) \right] \; ,
\end{eqnarray}
\end{widetext}
where 
\begin{equation}
A(1,2,4,5) =  \psi_h(1,2,3) f_\lambda({\cal M}^2_{45}) \; ,
\end{equation}
\begin{equation}
B(1,2,4,5) =  \psi_h(5,4,3) f_\lambda({\cal M}^2_{12}) \; ,
\end{equation}
and
\begin{equation}
T(1,2,4,5) = 
Tr\left[ S^\dagger(1,4) S(1,2) S^\dagger(5,2)S(5,4)\right] \; ,
\end{equation}
with $S(i,j) = \bar u_i v_j$, or 
\begin{equation}
S(i,j) = 
{\left[ m_q(x_i - x_j)\sigma^3 + (x_j k_i^\perp - x_i k_j^\perp )
\sigma^\perp\right] i \sigma^2 \over \sqrt{x_i x_j} } \; ,
\end{equation}
which give $T(1,2,4,5)$ equal to
\begin{equation}
2\, { (\vec y_1\vec y_2)(\vec y_3\vec y_4)   
     -(\vec y_1\vec y_3)(\vec y_2\vec y_4)  
     +(\vec y_1\vec y_4)(\vec y_2\vec y_3) 
       \over x_1 x_2 x_3 x_4}  \; , 
\end{equation}
where
\begin{eqnarray}
\vec y_1 &=&\left[m_q(x_1 - x_4), x_4 k_1^\perp - x_1 k_4^\perp\right]\; , 
\nonumber \\
\vec y_2 &=&\left[m_q(x_1 - x_2), x_2 k_1^\perp - x_1 k_2^\perp\right]\; , 
\nonumber \\
\vec y_3 &=&\left[m_q(x_5 - x_2), x_2 k_5^\perp - x_5 k_2^\perp\right]\; ,
\nonumber \\
\vec y_4 &=&\left[m_q(x_5 - x_4), x_4 k_5^\perp - x_5 k_4^\perp\right]\; .
\end{eqnarray}
An alternative expression is
\begin{eqnarray}
T(1,2,4,5) & = &  ({\cal M}^2_{14} - 4m_q^2)({\cal M}^2_{52} - 4m_q^2) \nonumber \\ 
           & - &  ({\cal M}^2_{15} - 4m_q^2)({\cal M}^2_{42} - 4m_q^2) \nonumber \\
           & + &  ({\cal M}^2_{12} - 4m_q^2)({\cal M}^2_{45} - 4m_q^2) \; .
\end{eqnarray}
\vskip.1in
In both parts of the amplitude, $A$ and $B$, one 
has $x_p = p^+/h^+$, $x_b = b^+/h^+ = 1- x_p$.
In meson $p$, one has $\vec k_{14} \equiv \vec k_p$, 
so that ${\cal M}_{14}={\cal M}_p=2 \sqrt{m_q^2 + \vec k_p^{\, 2}}$,
and the following relations hold: 
$x_{14} = (\sqrt{m_q^2 + \vec k^{\, 2}_p} + k_p^3)/ {\cal M}_p$,
$k_1^+ = x_{14} p^+$, 
$x_1 = x_{14} x_p$, 
$k_4^+ = (1-x_{14}) p^+$, 
$x_4 = (1-x_{14}) x_p$,
$k_1^\perp = x_{14} p^\perp + k_p^\perp$, 
$k_4^\perp = (1-x_{14}) p^\perp - k_p^\perp$.
In meson $b$, one has $\vec k_{52} \equiv \vec k_b$, 
so that ${\cal M}_b = 2\sqrt{m_q^2 + \vec k_b^{\, 2}} = {\cal M}_{52}$,
and the analogous relations are: 
$x_{52} = (\sqrt{m_q^2 + \vec k^{\, 2}_b} + k_b^3)/ {\cal M}_b$,
$k_5^+ = x_{52} b^+$, 
$x_5 = x_{52} x_b$, 
$k_2^+ = (1-x_{52}) b^+$, 
$x_2 = (1-x_{52}) x_b$,
$k_5^\perp = x_{52} b^\perp + k_b^\perp$, 
$k_2^\perp = (1-x_{52}) b^\perp - k_b^\perp$.
Evaluating the quarks' invariant masses in the hybrid and 
decay vertex, one then obtains
\begin{eqnarray}
{\cal M}^2_{12} &=& (x_1+x_2)
\left[{k_1^{\perp \, 2} + m^2_q \over x_1} 
+ {k_2^{\perp \, 2} + m^2_q \over x_2}\right] 
\nonumber \\
&-& (k_1^\perp + k_2^\perp)^2,
\end{eqnarray}
and
\begin{eqnarray}
{\cal M}^2_{54} &=& (x_5+x_4)
\left[{k_5^{\perp \, 2} + m^2_q \over x_5} 
+ {k_4^{\perp \, 2} + m^2_q \over x_4}\right] 
\nonumber \\
&-& (k_5^\perp + k_4^\perp)^2. 
\end{eqnarray}
In the part $A$ of the decay amplitude, one has 
\begin{equation}
\vec k^{\, 2}_q = {\cal M}^2_{12}/4 - m_q^2 \; ,
\end{equation}
and
\begin{equation}
\vec k^{\, 2}_g = 
{\left[{\cal M}^2_{123} - \left({\cal M}_{12} + m_g \right)^2\right]   
 \left[{\cal M}^2_{123} - \left({\cal M}_{12} - m_g \right)^2\right]
 \over
 4{\cal M}^2_{123} } \; ,
\end{equation}
where
\begin{eqnarray}
{\cal M}^2_{123} 
&=&
{k_1^{\perp \, 2} + m^2_q \over x_1} + 
{k_2^{\perp \, 2} + m^2_q \over x_2} 
\nonumber \\
&+& 
{(k_1 + k_2)^{\perp \, 2} + m^2_g \over 1 - x_1 - x_2}  
- (k_1 + k_2 + k_3)^{\perp \, 2} \; . 
\nonumber \\
&&
\end{eqnarray}
Similarly, in the part $B$, one has
\begin{equation}
\vec k^{\, 2}_q = {\cal M}^2_{54}/4 - m_q^2 \; ,
\end{equation}
and
\begin{equation}
\vec k^{\, 2}_g = 
{\left[{\cal M}^2_{543} - \left({\cal M}_{54} + m_g \right)^2\right]   
 \left[{\cal M}^2_{543} - \left({\cal M}_{54} - m_g \right)^2\right]
 \over 
 4{\cal M}^2_{543} } \; ,
\end{equation}
where
\begin{eqnarray}
{\cal M}^2_{543}  
&=&
{k_5^{\perp \, 2} + m^2_q \over x_5} + 
{k_4^{\perp \, 2} + m^2_q \over x_4} 
\nonumber \\
&+& 
{(k_5 + k_4)^{\perp \, 2} + m^2_g \over 1 - x_5 - x_4}  
- (k_5 + k_4 + k_3)^{\perp \, 2} \; .
\nonumber \\
&&
\end{eqnarray}


\begin{thebibliography}{99}
\bibitem{isgur}  
N. Isgur, R. Kokoski, J. Paton, 
Phys. Rev. Lett. {\bf 54} 869 (1985); 
F. Iddir, A. Le Yaouanc, L. Oliver, O. Pene, J.C. Raynal, 
Phys. Lett. B{\bf 207}, 325 (1988); 
P. R. Page, 
Phys. Lett. B{\bf 402}, 183 (1997).
\bibitem{lat1} 
K. J. Juge, J. Kuti, C. J. Morningstar,
Nucl. Phys. Proc. Suppl. {\bf 63}, 326 (1998);  
K. J. Juge, J. Kuti, C. J. Morningstar, 
Phys. Rev. Lett. {\bf 82}, 4400 (1999). 
\bibitem{CG} 
A. P. Szczepaniak, E. S. Swanson, 
Phys. Rev. D{\bf 55}, 3987 (1997);  
E. S. Swanson, A. P. Szczepaniak, 
Phys. Rev. D{\bf 59}, 014035 (1999);  
A. P. Szczepaniak, E. S. Swanson, 
Phys. Rev. D{\bf 65}, 025012 (2002).
\bibitem{PP1} 
T. Barnes, F. E. Close, F. de Viron, J. Weyers,
Nucl. Phys. B{\bf 224}, 241 (1983); 
M. S. Chanowitz, S. R. Sharpe,
Nucl. Phys. B{\bf 222}, 211 (1983);  
P. Hasenfratz, R. R. Horgan, J. Kuti, J. M. Richard, 
Phys. Lett. B{\bf 95}, 299 (1980); 
D. Horn, J. Mandula, 
Phys. Rev. D{\bf 17}, 898 (1978); 
N. Isgur, J. Paton, 
Phys. Rev. D{\bf 31}, 2910 (1985); 
A. W. Thomas, A. P. Szczepaniak, 
Phys. Lett. B{\bf 526}, 72 (2002). 
\bibitem{PP2} 
A. Le Yaouanc, L. Oliver, O. Pene, J. C. Raynal, S. Ono, 
Z. Phys. C{\bf 28}, 309 (1985); 
R. Kokoski, N. Isgur, 
Phys. Rev. D{\bf 35}, 907 (1987); 
F. E. Close, P. R. Page, 
Nucl. Phys. B{\bf 443}, 233 (1995);  
E. S. Swanson, A. P. Szczepaniak, 
Phys. Rev. D{\bf 56}, 5692 (1997); 
P. R. Page, E. S. Swanson, A. P. Szczepaniak,  
Phys. Rev. D{\bf 59}, 034016 (1999). 
\bibitem{LB}   
G. P. Lepage, S. J. Brodsky, 
Phys. Rev.  D{\bf 22}, 2157 (1980).
\bibitem{W6}   
K. G. Wilson et al., 
Phys. Rev. D{\bf 49}, 6720 (1994); 
and references therein.
\bibitem{GQCD} 
St. D. G{\l}azek, 
Phys. Rev. D{\bf 63}, 116006 (2001);
and references therein.
\bibitem{SGTM} 
St. D. G{\l}azek, T. Mas{\l}owski, 
Phys. Rev. D{\bf 65}, 065011 (2002).
\bibitem{SGMW} 
See e.g. St. D. G{\l}azek, M. Wi\c eckowski, 
hep-th/0204171.
\bibitem{RQM} 
A. P. Szczepaniak, C. R. Ji, S. R. Cotanch, 
Phys. Rev. D{\bf 52}, 5284 (1995);   
{\it ibid.} Phys. Rev. {\bf C}52, 2738 (1995).  
\bibitem{CG2}  
J. R. Finger, J. E. Mandula, 
Nucl. Phys. B{\bf 199}, 168 (1982); 
S. L. Adler, A. C. Davis, 
Nucl. Phys. B{\bf 244}, 469 (1984); 
A. Le Yaouanc, {\it et al.}, 
Phys. Rev. D{\bf 31}, 137 (1985);  
R. Alkofer, P. A. Amundsen, 
Nucl. Phys. B{\bf 306}, 305 (1988). 
A. P. Szczepaniak, E. S. Swanson, 
Phys. Rev. D{\bf 55}, 1578 (1997)
\bibitem{KS}   
See e.g. J. Kogut, L. Susskind, 
Phys. Rep. C{\bf 8}, 75 (1973).
\bibitem{Isgur1} 
S. Godfrey, N. Isgur, 
Phys. Rev. D{\bf 32}, 189 (1985); 
S. Capstick, N. Isgur, 
Phys. Rev. D{\bf 34}, 2809 (1986). 
\bibitem{Lor} 
B. L. G. Bakker, L. A. Kondratyuk, M. V. Terent'ev, 
Nucl. Phys. B{\bf 158} (1979) 497; 
L. A. Kondratyuk, M. V. Terentev, 
Sov. J. Nucl. Phys. {\bf 31}, 561 (1980). 
F. Coester, W. N. Polyzou, 
Phys. Rev. D{\bf 26}, 1348 (1982). 
\bibitem{DN}
See {\it e.g.} P. Danielewicz, J. M. Namys{\l}owski,
Phys. Lett. B{\bf 81}, 110 (1979).
\bibitem{vegas} 
G. P. Lepage, 
{\it Vegas: An adaptive multidimensional integration program},  
CLNS-80/447, March 1980. 
\bibitem{Powell}
W. H. Press, B. P. Flannery, S. A. Teukolsky, and W. T. Vettering,
{\it Numerical Recipes, The Art of Scientific Computing}
(Cambridge University Press, New York, 1986).
\bibitem{AP} 
B. H. Allen, R. J. Perry, 
Phys. Rev. D{\bf 62}, 025005 (2000).
\end{thebibliography}
\end{document}